\documentclass[twocolumn,showpacs,preprintnumbers,amsmath,amssymb]{revtex4}

\usepackage{graphicx}
\usepackage{dcolumn}
\usepackage{bm}
\newcommand{\bra}{\langle}
\newcommand{\ket}{\rangle}

\begin{document}

\title{Low--Energy Inelastic Neutrino Reactions on $^4$He}

\author{Doron Gazit}
\email{gdoron@phys.huji.ac.il}
\affiliation{The Racah Institute of Physics, The Hebrew University,
91904 Jerusalem, Israel.}
\author{Nir Barnea}
\email{nir@phys.huji.ac.il}
\affiliation{The Racah Institute of
Physics, The Hebrew University, 91904 Jerusalem, Israel.}

\date{\today}

\begin{abstract}
The inelastic scattering of neutrino off $^{4}$He is calculated
microscopically at energies typical for core collapse supernova
environment. The calculation is carried out with the Argonne V18
nucleon--nucleon potential and the Urbana IX three nucleon force. Full
final state interaction is included via the Lorentz integral
transform (LIT) method. The contribution of axial meson exchange currents to
the cross sections is taken into account from effective field
theory of nucleons and pions to order ${\cal{O}}(Q^3)$.
\end{abstract}


\maketitle

The current theory of core collapse supernova holds some open
questions regarding the explosion mechanism and late stage
nucleosynthesis. In order to analyze these questions, a better
understanding of the involved microscopical processes is needed. In
particular, due to the high abundance of $\alpha$ particles in
the supernova environment, the inelastic neutrino--$^4$He reaction
has drawn attention in recent years. This interest yielded a number
of studies trying to estimate the cross-section and the role of
neutrino--$^4$He reactions in the described phenomena,
\cite{HA88a,HA88b,GA04,YO05,YO06,OH06,SU06,WO90}. However to date, a
full {\it ab--initio} calculation that includes a realistic nuclear
Hamiltonian is still missing. Moreover, the contribution of meson
exchange currents (MEC) to this particular scattering process was
never estimated.

In this letter we present a full {\it{ab--initio}} calculation of
the inelastic neutrino--$^4$He reactions that meets these
challenges. Specifically, we consider the energy dependent inclusive
inelastic cross--sections for the following channels
$^4$He($\nu_x$,$\nu^{\,\prime}_x$)$^4_2$X,
$^4$He($\bar{\nu}_x$,$\bar{\nu}^{\,\prime}_x$)$^4_2$X,
$^4$He($\bar\nu_e$,e$^+$)$^4_1$X, and $^4$He($\nu_e$,e$^-$)$^4_3$X,
 where $x=e, \mu, \tau$ and $^A_Z$X stands for the final state $A$--nucleon
 system, with charge $Z$.

Core collapse supernovae are believed to be neutrino driven
explosion of massive stars. As the iron core of the star becomes
gravitationally
unstable it collapses until the nuclear forces halt the
collapse and drive an outgoing shock. This shock gradually stalls
due to energy loss
through neutrino radiation and dissociation of the iron nuclei into a mixture
of $\alpha$ particles and free nucleons.

At this stage, the proto-neutron star cools mainly by emitting neutrinos
in enormous numbers.
These neutrinos are a result of thermal pair production, and thus
are produced in flavor equilibrium. The characteristic temperatures of the
emitted neutrinos are
about $6-10$ MeV for $\nu_{\mu,\tau}$
($\bar{\nu}_{\mu,\tau}$), $5-8$ MeV for $\bar{\nu}_e$, and $3-5$ MeV for
$\nu_e$. The difference in temperature originates from the large cross-sections
for $\nu_e, \bar \nu_e$
electron scattering and charge current reactions.

At these temperatures there is a considerable amount of $\mu$ and
$\tau$ neutrinos (and anti-neutrinos) which carry more than $20$ MeV.
At such energies the $\nu_{\mu,\tau}$'s ($\bar \nu_{\mu,\tau}$'s)
 can dissociate the $^4$He
nucleus through inelastic neutral current reactions, thus creating
the seed to light element nucleosynthesis in the supernova
environment. A knock out of a nucleon from a $^4$He nucleus in the
helium rich layer, followed by a fusion of the remaining trinucleus
with another $\alpha$ particle, will result in a $7$--body nucleus.
This process is an important source of $^7$Li, and of $^{11}$B and
$^{19}$F through additional $\alpha$ capture reactions. Due to the
high dissociation energy of the $\alpha$, this mechanism is
sensitive to the high--energy tail of the neutrinos. Thus a correct
description of the process must contain an exact, energy dependent
cross-section for the neutral inelastic $\alpha-\nu$ reaction, which
initiates the process.

The relatively low temperature of the $\nu_e$'s and $\bar \nu_e$'s
emitted from the star's core suppress the probability for inelastic
reactions of these neutrinos with $^4$He in the supernova scenario.
Oscillations of the $\mu$ and $\tau$ (anti) neutrinos can yield a
secondary source of energetic electron neutrinos. Apparently, there
is a 1-3 neutrino oscillation resonance in O/C layer. As a result,
substantial increment in the charged current reaction rates, and
thus in the abundance of $^7$Li and $^{11}$B, can take place in the
helium layer, depending on the value of the mixing angle
$\theta_{13}$. In fact, such an effect can be used to constrain the
value of this oscillation parameter \cite{YO06}.

The possible role of inelastic $\nu-\alpha$ reactions in reviving
the supernova explosion shock was pointed out by Haxton
\cite{HA88a}. The hot dilute gas above the proto-neutron star and
below the accretion shock contains up to $70\%$ $^4$He nuclei. It is
believed that neutrinos emitted from the collapsed core (the
proto-neutron star) deposit energy in the matter behind the shock,
and eventually reverse the flow and revive the shock. This delayed
shock mechanism, originally introduced by Colgate and White
\cite{CO66}, has not yet been proved in full hydro-reactive simulations.
Haxton has suggested that inelastic neutral reactions of neutrinos
with $^4$He can lead to an enhanced neutrino energy deposition. This
effect is usually ignored (see however \cite{OH06,WO90}) and was not
yet considered in a full hydrodynamic simulation.

The first challenge in the study of the inelastic neutrino-$^4$He
reactions is the solution of the four--body problem, for ground and
excited states. As the $^4$He has no bound excited states, a
detailed knowledge of the four nucleons continuum is needed to
assure the final state interaction (FSI). This makes an explicit
calculation impossible, since a complete description of the nuclear
four--body system is currently out of reach. We avoid this
complication by calculating the FSI through the Lorentz integral
transform (LIT) method \cite{LIT1}. For the solution of the ground
state wave function and the LIT equations we use the effective
interaction in the hyperspherical harmonics (EIHH) approach
\cite{EIHH1,EIHH2,symhh}. For the nuclear Hamiltonian, we take the
nucleon-nucleon (NN) potential Argonne V18 (AV18) \cite{AV18} with
the Urbana IX (UIX) \cite{UIX} three nucleon force (3NF). This
Hamiltonian has been used successfully to reproduce the spectra of
light nuclei \cite{UIX}, and electro-weak reactions with light
nuclei \cite{GA06,MA01,SC98}.

In the limit of small momentum transfer with respect to the mass of the
$Z,W^{\pm}$ bosons, the weak interaction Hamiltonian is given by
 $\hat{H}_{W}=-\frac{G}{\sqrt{2}}\int {d^{3}x
\hat{j}_{\mu }(\vec{x}) \hat{J}^{\mu }(\vec{x})}$, where $G$ is the
Fermi weak coupling constant, $\hat{j}_{\mu }(\vec{x})$ is the
lepton current, and $\hat{J}^{\mu }$ is the nuclear current. The
lepton is a point Dirac particle, and evaluating its current and its
contribution to the cross-section is relatively simple, yielding
only kinematical factors.
 The nuclear current, however, is more complicated. The formal
structure of the nuclear weak neutral current is,
\begin{equation}\label{eq:NC}
\hat{J}_{\mu }^0=(1-2\cdot \sin ^{2}\theta _{W})\frac{\tau
_{0}}{2}\hat{J}^V_{\mu }+\frac{\tau _{0}}{2}\hat{J}^A_{\mu}-2\cdot
\sin ^{2}\theta _{W}\frac{1}{2 }\hat{J}^V_{\mu } \;,
\end{equation}
and the structure of the charged currrents is,
\begin{equation}\label{eq:CC}
\hat{J}_{\mu }^{\pm}=\frac{\tau _{\pm}}{2}\hat{J}^V_{\mu
}+\frac{\tau _{\pm}}{2}\hat{J}^A_{\mu} \;,
\end{equation}
where the superscript $A$ ($V$) stands for axial (vector) currents.

The leading contributions to these operators are the
one--body terms. It is well known, however, that mesonic degrees of freedom
 can contribute to the nuclear currents through many--body terms,
namely MEC, even if they do not appear explicitly in the
Hamiltonian. The modern point of view \cite{WE90} has created a
systematic way of considering these degrees of freedom, that is the
effective field theory (EFT) approach. EFT is based on the idea that
an observable characterized by a momentum $Q$, does not depend on
momenta much higher than $Q$. One introduces a cutoff momentum
$\Lambda$, and integrates out the degrees of freedom present at $Q$ larger than
$\Lambda$. A perturbation theory in the small parameter $Q/\Lambda$
can now be developed systematically. The coefficients of the
different terms are called low--energy constants (LEC), usually
calibrated in experiments or by theory. EFT has two major
advantages, one is the link to the underlying high--energy theory,
which in the case of the strong interaction is commonly believed to
be Quantum Chromodynamics (QCD). The other advantage is the ability
to provide a control of the accuracy in the calculation. The problem
with EFT is that while a percentage level accuracy in describing
scattering process is already achieved using a next--to--leading order
(NLO) lagrangian, this is not the case when trying to recover
successfully the wealth of experimental data described by the
phenomenological approach (nuclear binding energies for example).
This task demands at least next--to--next--to--next--to--leading
order (N$^3$LO) EFT lagrangians \cite{EFT1,EFT2}. It is thus clear
that a hybrid method that joins together the success of the standard
nuclear physics approach and the clear advantages of EFT is called
for, although the resulting MEC will not be completely consistent
with the nuclear Hamiltonian. This hybrid approach was coined by Rho
as MEEFT ("more effective EFT") \cite{rho06}, and was already
applied to study electro-weak reactions for $A=2,3,4$ nuclei
\cite{PA03}. In this work we adopt the hybrid approach combining the
phenomenological AV18 and UIX nuclear potentials with EFT based
nuclear MEC.

The
conservation of vector current (CVC) hypothesis states that the
vector current is an isospin rotation of the electromagnetic
current. Thus, the electric part of the vector meson exchange currents can be approximated very well
at low q via the Siegert theorem, from the single nucleon vector charge operator.
That is not the case for the axial current, which is not conserved and should
be calculated explicitly.
For this task we shall use the EFT meson exchange currents.

The typical energy scale of the neutrino in the supernova
environment is some tens of MeV, thus a proper cutoff is of at least
a few hundred MeV. It is important to notice that the cutoff should
not be higher than the mass of the nucleon, which is the order of
the QCD mass scale. We will use cutoff values in the range
$\Lambda=400-800$ MeV. In the EFT scheme employed here, nucleons and
pions are the explicit degrees of freedom. The model includes the
pions as goldstone bosons of the chiral symmetry \cite{EFT3}. The
axial currents are the N\"{o}ther currents derived from a NLO order
lagrangian, in a relativistic approach. These currents are accurate
to N$^3$LO and are given in momentum space, as they originate from
a Lorentz invariant theory. For the transformation to configuration
space, we perform a
Fourier transform with cutoff \cite{PA03},
\begin{equation}
{\hat{O}(\vec{x})=\int \frac{d^{3}\vec{q}}{(2\pi)^3}
\hat{O}(\vec{q}) S_\Lambda(q)}.
\end{equation}
The cutoff function $S_\Lambda(q)$ is $1$ for $q\ll \Lambda$, and
approaches $0$ for $q\gg \Lambda$. We use a gaussian cutoff
function as proposed by Park \textit{et. al.} \cite{PA03}, i.e.
$S_\Lambda(q)=exp(-\frac{q^2}{\Lambda^2})$. It is important to note
that this method leads to the same single nucleon operators as the
standard nuclear physics approach. The meson exchange currents in
configuration space are the Fourier transform of propagators with a
cutoff, which in the limit $\Lambda \rightarrow \infty$ are just the
usual Yukawa functions. In contrast to the standard nuclear physics
approach, the coefficients of the functions are not structure
functions, but LECs. All the LECs which originate from a
nucleon--pion interaction, namely $g_A$, $f_\pi$, $c_3$, and $c_4$,
are calibrated using low--energy pion--nucleon scattering. Alas, in
this order there are also two nucleon contact terms, which introduce
LECs that can be calibrated only by nuclear matter processes.
Fortunately, using Lorentz invariance, the axial currents introduce
only one unknown LEC, which is denoted by $\hat{d}_r$. This
coefficient can be calibrated by the triton half life time for each value
of $\Lambda$, thus $\hat{d}_r=\hat{d}_r(\Lambda)$. In our
calculations we reproduce the results published by Park \textit{et.
al.} \cite{PA03}, for the cutoff dependency.

For low--energy reactions a multipole decomposition of the currents
is useful. Applying Fermi's golden rule, to inclusive reactions with unpolarized targets,
and considering recoil effects, the differential
cross-section takes the form \cite{DO76},
\begin{eqnarray}
\lefteqn{\left(\frac{d\sigma^a }{dk_{f}}\right)_{\nu
(\bar{\nu})}=\frac{2G^{2}}{2J_{i}+1} k_{f}^{2} F^a(Z_f,k_f)} \cr &&
{\int d\epsilon \int_{0}^{\pi } \sin {\theta }d\theta \delta
\left(\epsilon -\omega +\frac {q^{2}}{2M_{^{4}\mathrm{He}}}\right) }
\cr &&  {\left\{
    \sum_{J=0}^\infty \left[ X_{\hat{C}}
       {R}_{\hat{C}_J}  + X_{\hat{L}} {R}_{\hat{L}_J} - X_{\hat{C}\hat{L}} Re {{R}_{\hat{C}^*_J {\hat{L}}_J}}
        \right] \right.}
\cr & &
       {\left. + \sum_{J=1}^\infty \left[X_{\hat{M}}R_{\hat{M}_J} + X_{\hat{E}} R_ {\hat{E}_J}  \mp
       X_{\hat{E}\hat{M}} Re {{R}_{\hat{E}^*_J \hat{M}_J}} \right] \right\}}
       \;,
\end{eqnarray}
where ${k}_{f}$ is the momentum of the outgoing lepton, $J_i=0$ is
the angular momentum of the $^4$He, $Z_f$ is the charge of the
residual nuclear system. The four-vector $(\omega,\vec q)$
represents energy and momentum transfer, and $\theta $ is the angle
between the incoming neutrino direction and outgoing lepton
direction. The superscript $a$ denotes the isospin component, with
$a=0$ for the neutral current and $a=\pm$ for the charged currents.
The Coulomb factor $F^a(Z,k)$ is equal to $1$ for neutral currents,
and is the Fermi function for charged current. The functions
$X_{\hat{O}_1\hat{O}_2}$ are the leptonic kinematical factors
(related to the $\hat{O}_1$, $\hat{O}_2$ multipoles,
$X_{\hat{O}_1}=X_{\hat{O}_1\hat{O}_1}$). They depend on the mass and
the momentum of the outgoing lepton. Similarly, the functions
$R_{\hat{O}_{1}\hat{O}_{2}}(\epsilon,q)$
are the nuclear response functions.
The transition operators
$C_{J}(q), L_{J}(q), E_{J}(q), M_{J}(q)$ are the reduced Coulomb,
longitudinal, transverse electric and transverse magnetic operators
of angular momentum $J$. The response functions are calculated by
inverting the Lorentz integral transforms
\[
{L_{\hat{O}_{1}\hat{O}_{2}}(\sigma,q)= \int d\epsilon
\frac{R_{\hat{O}_{1}\hat{O}_{2}}(\epsilon,q )}{(\epsilon -\sigma
_{R})^{2}+\sigma _{I}^{2}}=\langle \tilde{\Psi}_{1}\mid \tilde{\Psi}
_{2}\rangle },
\]
where $\sigma =\sigma _{R}+i\sigma _{I}$, and $\mid\tilde{\Psi}_{i}\rangle $
($i=1,2$) are solutions of the Schr\"{o}dinger like equations
\[
{(H-E_{0}-\sigma )\mid \tilde{\Psi}_{i}(\sigma,q )\rangle
=\hat{O_{i}}(q)\mid \Psi _{0}\rangle }.
\]
The localized character of the ground state, and the imaginary part
of $ \sigma $, give these equations an asymptotic boundary condition
similar to a bound state. As a result, one can solve these equations
with the hyperspherical harmonics (HH) expansion using the EIHH
\cite{EIHH1,EIHH2,symhh} method. The matrix elements $\bra
\tilde{\Psi}_{1} | \tilde{\Psi}_{2}\ket$ are calculated with the
Lanczos algorithm \cite{MA02}.

\begin{figure}
\rotatebox{0}{ \resizebox{7.1cm}{!}{
\includegraphics{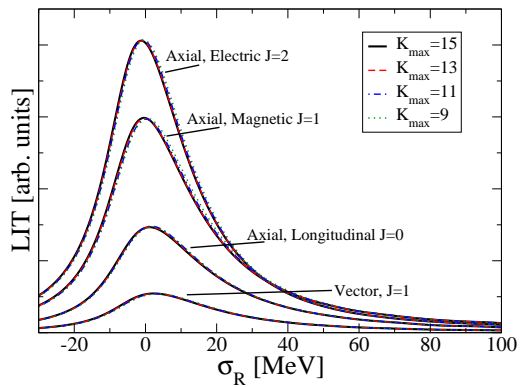}  } }
\caption{\label{fig:conv} (color online) LIT convergence for the
leading multipoles, as a function of the HH grand angular momenta K.
}
\end{figure}

In the supernova scenario one has to consider neutrinos with up to
about $60$ MeV. Usually, the leading contributions in weak nuclear
processes are the Gamow-Teller and the Fermi operators. Due to the
total angular momentum and spin structure of the $^{4}\mathrm{He}$
nucleus, they are both strongly suppressed. In fact, the
Gamow-Teller operator contributes only due to the small $P$-- and
$D$--wave components of the ground state wave function. The same
argument follows for the $M^V_1$ operator. In addition,
$^{4}\mathrm{He}$ is an almost pure zero--isospin state
\cite{NO02,VI05}, hence the Fermi operator vanishes. Therefore, the
leading contributions to the inelastic cross-section are due to the
axial vector operators $E^A_2, M^A_1, L^A_2, L^A_0 $ and the vector
operators $C^V_1, E^V_1, L^V_1$ (the latter are all proportional to
each other due to the Siegert theorem). For the neutrino energies
considered here it is sufficient to retain contributions up to
$O(q^2)$ in the multipole expansion \cite{GA04}. In Fig.
\ref{fig:conv} we present for these multipoles the convergence of
the LIT  as a function of the principle HH grand angular-momentum
quantum number $K$. It can be seen that the EIHH method results in a
rapid convergence of the LIT calculation to a sub-percentage
accuracy level. Comparing with a previous work, \cite{GA04}, we
conclude that the 3NF does not affect much the convergence rate of
these operators.

\begin{table}
\begin{tabular}{c||c|c|c|c}
\hline \hline T [MeV] &  \multicolumn{4}{c} {$\bra \sigma^0_x \ket_T
=  \frac{1}{2} \frac{1}{A} \bra \sigma_{\nu_x}^0+
\sigma_{\overline{\nu}_x}^0 \ket_T$ [$10^{-42}cm^{2}$] }  \\ \hline
 &  AV8' \cite{GA04} & AV18 & AV18+UIX & AV18+UIX+MEC \\
\hline
 4    &  2.09(-3) & 2.31(-3) & 1.63(-3) & 1.66(-3) \\
 6    &  3.84(-2) & 4.30(-2) & 3.17(-2) & 3.20(-2) \\
 8    &  2.25(-1) & 2.52(-1) & 1.91(-1) & 1.92(-1) \\
 10   &  7.85(-1) & 8.81(-1) & 6.77(-1) & 6.82(-1) \\
 12   &  2.05     & 2.29     & 1.79     & 1.80    \\
 14   &  4.45     & 4.53     & 3.91     & 3.93    \\
\hline \hline
\end{tabular}
\caption{{\label{tab:pots}} Temperature averaged neutral current
inclusive inelastic cross-section per nucleon (in $10^{-42}cm^{2}$)
as a function of neutrino temperature (in MeV). }
\end{table}
\begin{table}
\begin{tabular}{c||c|c|c|c}
\hline \hline T &  \multicolumn{4}{c}{$\langle \sigma \rangle_T$
[$10^{-42}cm^{2}$] }  \\ \hline
 [MeV] &  ($\nu_x$,$\nu^{\,\prime}_x$)             &
          ($\bar{\nu}_x$,$\bar{\nu}^{\,\prime}_x$) &
          ($\nu_e$,e$^-$)                      &
          ($\bar\nu_e$,e$^+$) \\
\hline
 2    & 1.47(-6)  & 1.36(-6) & 7.40(-6) & 5.98(-6) \\
 4    & 1.73(-3)  & 1.59(-3) & 8.60(-3) & 6.84(-3) \\
 6    & 3.34(-2)  & 3.07(-2) & 1.63(-1) & 1.30(-1) \\
 8    & 2.00(-1)  & 1.84(-1) & 9.61(-1) & 7.68(-1) \\
 10   & 7.09(-1)  & 6.54(-1) &  3.36    & 2.71    \\
 \hline \hline
\end{tabular}
\caption{{\label{tab:crs_t}} Temperature averaged inclusive
inelastic cross-section per nucleon as a function of temperature. }
\end{table}
\begin{table}
\begin{tabular}{c||c|c|c|c}
\hline \hline T &  \multicolumn{4}{c}{$\langle \sigma \omega
\rangle_T$ [$10^{-42}$MeV cm$^{2}$] }  \\ \hline
 [MeV] &  ($\nu_x$,$\nu^{\,\prime}_x$)             &
          ($\bar{\nu}_x$,$\bar{\nu}^{\,\prime}_x$) &
          ($\nu_e$,e$^-$)                      &
          ($\bar\nu_e$,e$^+$) \\
\hline
 2    &  3.49(-5) & 3.23(-5) &  1.76(-4)& 1.42(-4) \\
 4    &  4.50(-2) & 4.15(-2) &  2.27(-1)& 1.80(-1)\\
 6    &  9.26(-1) & 8.56(-1) &  4.56    & 3.70    \\
 8    &  5.85     & 5.43     &  28.4     & 22.9  \\
 10   &  21.7     & 20.2     &  103.8    &  84.4    \\
 \hline \hline
\end{tabular}
\caption{{\label{tab:crsomega_t}} Temperature averaged inclusive
inelastic energy transfer cross-section per nucleon as a function of
temperature. }
\end{table}
It is customary to assume that supernova neutrinos are in thermal
equilibrium, so their spectra can be approximated by the Fermi-Dirac
distribution with a characteristic temperature $T$. In Table
\ref{tab:pots} we present the temperature averaged total neutral
current inelastic cross--section as a function of the neutrino
temperature for the AV8', AV18, and the AV18+UIX nuclear
Hamiltonians and for the AV18+UIX Hamiltonian adding the MEC. From
the table it can be seen that the low--energy cross--section is
rather sensitive to details of the nuclear force model (the effect
of 3NF is about $30\%$). Sensitivity that gradually decreases with
growing energy. In contrast the effect of MEC is rather small in our
case, being on the percentage level. Similar tendencies were also
observed   for the {\it hep} process \cite{MA01}. The small
contribution of the MEC can be understood in the following way. The
symmetry of the vertices in this low--energy approximation, dictate
a symmetry between the two nucleons interacting via the meson
exchange. The leading one--body multipoles have negative parity, as
a result the MEC contributions to them is small. In comparison the
MEC correction to the Gamow-Teller is of the same magnitude as the
one--body current, however both terms become marginal with
increasing momentum transfer. Although presented for the neutral
current, these arguments hold true also for the charged currents
since the response functions are related by isospin rotation.

In Table \ref{tab:crs_t} and Table \ref{tab:crsomega_t} we present
(for AV18+UIX+MEC) the temperature averaged cross--section and
energy transfer as a function of the neutrino temperature for the
various processes. In both tables it can be seen that the charged
current process is roughly a factor five more efficient than the
neutral current process. Our results are of the same order of
magnitude of previous estimates by Woosley et. al. \cite{WO90},
though the differences can reach $25\%$. The current work predicts a
stronger temperature dependence, with substantial increment at high
temperatures. This indicates a different structure of the predicted
resonances.

Summarizing, we have given the first full microscopic study of
$\nu-\alpha$ reactions, using a state of the art nuclear Hamiltonian
including MEC.
The overall accuracy of our calculation is of the order of $5\%$.
This error is mainly due to the strong sensitivity of the
cross--section to the nuclear model, which could enhance the effect
of uncertainties in the Hamiltonian. The numerical accuracy of our
calculations is of the order of $1\%$, and the cutoff dependence of
the MEC is of the same order. With the present calculation, we make
an important step in the path towards a more robust and reliable
description of the neutrino heating of the pre--shock region in
core--collapse supernovae, in which $^4$He plays a decisive role.

We would like to thank G. Orlandini and W. Leidemann for their
useful comments. This work was supported by the ISRAEL SCIENCE FOUNDATION
(grant no.~361/05).

\bibliography{nu_alpha_v2}
\end{document}